\def\a{\alpha}\def\d{\delta}\def\e{\epsilon}
\def\f{\phi}\def\g{\gamma}\def\h{\theta}
\def\l{\lambda}\def\m{\mu}\def\n{\nu}\def
\p{\pi}\def\q{\psi}\def\r{\rho}\def\s{\sigma}
\def\y{\eta}

\def\D{\Delta}
\def\O{\Omega}\def\S{\Sigma}

\def\de{\partial}\def\na{\nabla}\def\lra{\leftrightarrow}
\def\inf{\infty}\def\id{\equiv}\def\mo{{-1}}
\def\({\left(}\def\){\right)}\def\[{\left[}\def\]{\right]}
\def\mn{{\mu\nu}}

\def\fe{field equations }\def\bh{black hole }
\def\coo{coordinates }

\def\cc{coupling constant }

\def\sch{Schwarzschild }
\def\RN{Reissner-Nordstr\"om }

\def\ab{asymptotic behavior }
\def\ie{i.e.\ }

\def\bc{boundary conditions }

\def\section#1{\bigskip\noindent{\bf#1}\smallskip}

\def\subsection#1{\smallskip\noindent{\it#1}\smallskip}

\font\smalll = cmr7

\def\PL#1{Phys.\ Lett.\ {\bf#1}}
\def\PRL#1{Phys.\ Rev.\ Lett.\ {\bf#1}}
\def\PR#1{Phys.\ Rev.\ {\bf#1}}
\def\NP#1{Nucl.\ Phys.\ {\bf#1}}\def\GRG#1{Gen.\ Relativ.\ Grav.\ {\bf#1}}
\def\JMP#1{J.\ Math.\ Phys.\ {\bf#1}}

\def\JoP#1{J.\ Phys.\ {\bf#1}} \def\IJMP#1{Int.\ J. Mod.\ Phys.\ {\bf #1}}
\def\MPL#1{Mod.\ Phys.\ Lett.\ {\bf #1}} 
\def\PRep#1{Phys.\ Rep.\ {\bf#1}}
\def\AoP#1{Ann.\ Phys.\ {\bf#1}}

\def\arx#1{{\tt arXiv:#1}}
\def\PRep#1{Phys.\ Rep.\ {\bf#1}}\def\IJTP#1{Int.\ J. Theor.\ Phys.\ {\bf #1}}

\def\ref#1{\medskip\everypar={\hangindent 2\parindent}#1}
\def\beginref{\begingroup
\bigskip
\centerline{\bf References}
\nobreak\noindent}
\def\endref{\par\endgroup}

\input epsf
\def\cL{{\cal L}}\def\cR{{\cal R}}\def\cS{{\cal S}}
\def\ug{\;=\;}

{
\line{}
\vskip30pt
\centerline{\bf Dyonic black holes from dimensional reduction}
\smallskip
\centerline{\bf of five-dimensional Einstein-Gauss-Bonnet gravity}

\vskip60pt
\centerline{
{\bf Z. Belkhadria} $^1$\footnote{$^\star$}{e-mail: zakaria.belkhadria@gmail.com},
{\bf S. Mignemi} $^{1,2}$\footnote{$^\ddagger$}{e-mail: smignemi@unica.it} and {\bf F. Paderi} $^3$\footnote{$^\dagger$}{e-mail: fedpad99@gmail.com}}

\vskip10pt

\smallskip
\centerline{$^1$Dipartimento di Matematica e Informatica, Universit\`a di Cagliari,}
\centerline{via Ospedale 72, 09124 Cagliari, Italy}
\smallskip
\centerline{$^2$INFN, Sezione di Cagliari, Cittadella Universitaria, 09042 Monserrato, Italy}
\smallskip
\centerline{$^3$Dipartimento di Fisica, Universit\`a di Cagliari,}
\centerline{Cittadella Universitaria, 09042 Monserrato, Italy}

\vskip80pt

\centerline{\bf Abstract}
\medskip
{\noindent We study the general black hole solutions of dimensionally reduced five-dimensional Einstein-Gauss-Bonnet gravity.
The reduced theory contains gravity, electromagnetism and a scalar field, with nonlinear corrections to the action and nontrivial couplings.
The solutions can be classified through mass and electric and magnetic charge. They  present peculiar features
with respect to the solutions of the standard Kaluza-Klein theory without Gauss-Bonnet corrections, like the existence of an extremal mass
even in the neutral case.
Also the thermodynamics is affected, for example, extremal black holes display nonvanishing temperature and entropy.
\vskip10pt
{\noindent

}
\vfill\eject\

}

\section{1. Introduction}
Soon after the discovery of general relativity, Kaluza proposed a theory unifying gravity and electromagnetism
based on the hypothesis that spacetime has five dimensions and its dynamics is governed by the Einstein-Hilbert
Lagrangian [1].
The theory was then improved by Klein, who observed that the fifth dimension would not be observable if it were curled
up in an exceedingly small circle [2].

The theory of Kaluza included only gravity and electromagnetism, but later it was observed that also a scalar field,
sometimes called dilaton  with a term borrowed from string theory,
can arise naturally from the formalism [3]. Much later, it was also argued that the introduction of the dilaton is
necessary for the consistency of the field equations [4].

The Kaluza-Klein (KK) theory had a revival in the 80's.
Around that time, it was observed that in higher dimensions the Einstein-Hilbert Lagrangian is no longer unique, but some
higher-derivative terms -- called Lovelock or Gauss-Bonnet (GB) terms -- can be added to it, still leading to second order
field equations [5-7].
In lower dimensions, these terms become total derivatives and do not affect the field equations.
Their dimensional reduction can however give rise to correction both to the gravitational and the Maxwell Lagrangian.
More specifically,  the Maxwell Lagrangian acquires terms that give rise to a specific form of nonlinear electrodynamics [8],
while the gravitational Lagrangian acquires a term quadratic in the Riemann tensor, containing a nonminimal coupling
to the dilaton that contributes to the field equations only for nontrivial scalar field.
In addition, nontrivial couplings of Maxwell and Riemann terms arise [9]. The action so obtained still gives rise to second-order
\fe and is therefore of Horndeski type [10].

The GB correction to the higher-dimensional Lagrangian can also arise from string theory [11]: in this context it was shown
that it does not add new degrees of freedom to the theory besides the graviton, and therefore the spectrum of particles is
free from ghosts or tachyons.

The discussion of black holes in theories containing nonminimal coupling of gravity and electromagnetism to a dilatonic
field are important for uniqueness and no hair theorems [12]. These theorems were stated in the 70's and showed that the
Einstein-Maxwell theory minimally coupled to a scalar field admits only solutions of Kerr-Newman type, parametrized
by mass, angular momentum and  electric  and magnetic charges, that reduce to the \RN (RN) metric in the case of spherical symmetry.
Some counterexamples were found later when additional fields were included in the theory,
like skyrmions [13] or gauge fields [14], or nonminimal couplings of the Maxwell field to a scalar [15-17].
Also if the GB term is coupled nonminimally to the dilaton, solutions with nontrivial scalar field can be found [18-21].
In this case, the occurrence of a minimal admissible value for the mass was observed.
Usually, in these solutions the dilaton charge is not an independent parameter, but is a function of the other conserved charges
and has therefore been called secondary hair [22].

Another interesting example of black holes violating the no-hair theorems are dyonic (\ie possessing both electric and magnetic
charge) solutions of the standard KK theory. These have been originally obtained by Dobiasch and Maison (DM) [23], and are
discussed in several papers [24-26]. They show peculiar
features, as a duality symmetry between electric and magnetic solutions. In general, they possess two horizon, but reduce to
the solutions of refs.~[15-17] with a single horizon when one of the charges vanishes.
Although magnetic charges are not observed in nature these solutions are particularly interesting from a theoretical point of
view.

In previous work [27,28], we have investigated dyonic black hole solutions of higher-derivative (GB) KK theory when the
scalar field is neglected, showing the existence of nontrivial solutions also in that case. These solutions do not respect
the electric-magnetic duality because of the presence of the GB interaction. Depending on the range of the parameters, they can
possess one or two horizons. In some cases, they support an everywhere regular electric field.\footnote{$^1$}{Solutions of this theory,
possessing only electric charge, have also been studied in [29] in a five-dimensional setting.}

In this paper we expand the investigations started in [27,28] by introducing a scalar field in the ansatz of dimensional
reduction of the higher-derivative KK theory. The solutions will give corrections to the \bh solution of standard KK [23-26].
In general, they do not present major differences, except that for a range of values of the charges, the inner horizon disappears.
In fact, the corrections are effective only for small values of the radius, while the asymptotic behaviour remains identical.
Also, for some values of the parameters, the electric field is regular at the singularity. Moreover, a minimal value of the mass
is present also in the case of vanishing charges, like in pure dilatonic Einstein-GB \bh [23-26].

We also discuss the thermodynamics of these solutions. Now the GB corrections modify the small-mass behaviour, which however remains
qualitatively similar to that of the DM solutions.
If $Q=0$ or $P=0$, the temperature has a maximum at the extremal value $M_{ext}$ of the mass, otherwise it increases from  a finite value
at $M_{ext}$ toward a maximum and then decreases to 0 for $M\to\inf$.

\section{2. The action}
We consider the five-dimensional action
$$I^{(5)}=\int d^5x\sqrt{-g^{(5)}}\[\cR^{(5)}+{\a\over4}\,\cS^{(5)}\],\eqno(2.1)$$
where $\a$ is a \cc with dimension $[L]^2$ and $\cS=\cR^{\m\n\r\s}\cR_{\m\n\r\s}-4\cR^{\m\n}\cR_{\m\n}+\cR^2$ is the Gauss-Bonnet term,
and make the standard ansatz
$$g_{AB}=\(\matrix{e^\f g_\mn+4e^{-2\f}A_\m A_\n&2e^{-2\f}A_\m\cr2e^{-2\f}A_\n&e^{-2\f}}\),\eqno(2.2)$$
where $g_\mn$ is the four-dimensional metric, $A_\m$ is the Maxwell 4-potential and $\f$ is a scalar field.
After dimensional reduction, the action becomes
$$I=\int d^4x\,\cL,\eqno(2.3)$$
where the four-dimensional Lagrangian density is [9]
$$\eqalignno{\cL&=\sqrt{-g}\Big[\cR-6(\na\f)^2-e^{-6\f}F^2\cr&+\a\big[e^{-2\f}\cS+3e^{-14\f}\big((F_\mn F^\mn)^2-F^\mn F_{\n\r}F^{\r\s}F_{\s\m}\big)
+16e^{-2\f}G_\mn\na^\m\f\na^\n\f \cr&-2e^{-8\f}(R_{\m\n\r\s}-4R_{\m\r}g_{\n\s}+Rg_{\m\r} g_{\n\s})F^\mn F^{\r\s}+4e^{-8\f}F^\mn(2\na_\m\f\na^\r F_{\n\r}
-\na^\r\f\na_\r F_\mn)\cr&+8e^{-8\f}\big(4F^\mn F_\mn(\na\f)^2-7F_\mn F^{\m\r}\na^\n\f\na_\r\f\big)+24e^{-2\f}(\na\f)^2\na^2\f\Big].&(2.4)}$$
$F_\mn=\de_\m A_\n-\de_\n A_\m$ is the Maxwell field strength and $\cR$ and $\cS$ are the four-dimensional Ricci  scalar and GB term.
We have adopted a non-standard normalization for the scalar field.
For $\a=0$ we recover the Einstein action nonminimally coupled to the Maxwell field by a scalar, as in the standard KK theory.

We look for dyonic solutions with spherical symmetry.  Therefore, we make the ansatz
$$ds^2=-\D(r)dt^2+{dr^2\over\s^2(r)\D(r)}+R^2(r)d\O^2,\qquad\f=\f(r),\eqno(2.5)$$
$$A=A_0(r)\,dt+ P\cos\h\,d\f,\eqno(2.6)$$
with $A_0(r)$ the electric potential and $P$ the magnetic charge. We then substitute the ansatz into the action:
the integration over the angular variables is trivial and
after several integration by parts on the variable $r$, one gets, up to a multiplicative constant,
$$\eqalignno{\cL&=\s^\mo\[1-{P^2\over R^2}e^{-6\f}\]+\s\bigg[\D'R'R+\D R'^2-3\D R^2\f'^2+A_0'^2R^2e^{-6\f}\bigg]+\a\s\bigg[-3{P^2A_0'^2\over R^2}e^{-14\f}\cr
&-(\D'\f'+2\D\f'^2)e^{-2\f}+\(A_0'^2+{P^2\over R^3}\big(\D'R'+4\D'R\f'+8\D R\f'^2+2\D R'\f'\big)\)e^{-8\f}\bigg]\cr
&+\a\s^3\D\bigg[\(\D'R'^2\f'+2\D'RR'\f'^2+2\D R'^2\f'^2+\D'R^2\f'^3+4\D RR'\f'^3+2\D^2R^2\f'^4\)e^{-2\f}\cr&-A_0'^2(R'^2+R^2\f'^2)e^{-8\f}\bigg],
&(2.7)}$$
where a prime denotes a derivative with respect to $r$.
From this expression, one can obtain the field equations by performing a variation with respect to the fields $\D$, $\s$,$R$, $\f$ and $A_0$.
Only four of the ensuing equations are independent, so a gauge can be fixed by constraining one of the metric functions.

\section{3. Special case $\a=0$}
Before investigating the general solutions of (2.7), we recall that for $\a=0$ an exact solution of the field equations can be found [23,25,26].
In that case, one has an electric-magnetic duality for $\f\lra-\f$, $e^{-6\f}F_\mn\lra {^*\!F_\mn}$, that is spoiled when $\a\ne0$.
The solutions can be obtained in closed form by choosing the gauge $\s=1$, hence taking a metric
$$ds^2=-\D(r)dt^2+{dr^2\over\D(r)}+R^2(r)d\O^2,\qquad\f=\f(r),\eqno(3.1)$$
and depend on three parameters $a$, $b$, and $c$. They are given by [23]
$$\D={f^2\over gh},\qquad R^2=gh,\qquad e^{2\f}={g\over h},\qquad E=A_0'=Q\ {g^2\over h^4},\eqno(3.2)$$
where $E$ is the electric field and
$$f^2=\(r-{a+b\over4}\)^2-{c^2\over4},$$
$$g^2=\(r+{a-b\over4}\)^2-{(a^2-c^2)(a-b)\over4(a+b)},\qquad h^2=\(r-{a-b\over4}\)^2-{(b^2-c^2)(b-a)\over4(a+b)}.\eqno(3.3)$$

The parameters are related to the physical mass $M$, dilatonic charge $D$, and electric and magnetic charges $Q$ and $P$ by
$$M={a+b\over4},\qquad D={b-a\over4},\qquad Q^2={b(b^2-c^2)\over4(a+b)}, \qquad P^2={a(a^2-c^2)\over4(a+b)},\eqno(3.4)$$
where we have defined $D$  such that $\f\sim-{D\over r}$ at infinity. The physical parameters satisfy the constraint
$${Q^2\over D+M}+{P^2\over D-M}=2D,\eqno(3.5)$$
so that only three of them are independent. The scalar charge can then be interpreted as a secondary hair (see [22]).
The electric-magnetic duality implies that the solutions are invariant for $Q\lra P$, $D\lra-D$.

In general, these solutions are singular at the zeroes of $g$ and $h$ and present horizons at the zeroes of $f$, namely at $r_\pm=(a+b\pm2c)/4$.
Assuming positive $a$, the solutions have two horizons if $a>c$ and $b>c$.  The limits $b=c$, $a=c$ correspond
to black holes with vanishing electric and magnetic charge, respectively and coincide with the solutions found in [15-17].
In these cases, a single horizon occurs.
For $a=b\ne c$, one recovers the \RN solution, and if $a=b=c$, the \sch metric.
Black holes become  extremal if $c=0$, \ie $M^2+3D^2=Q^2+P^2$: it follows that $M_{ext}^2\le Q^2+P^2$.

The temperature and the entropy of the black hole can be obtained from standard formulae and read
$$T={1\over2\p}{c(a+b)\over\sqrt{ab}\,(a+c)(b+c)},\eqno(3.6)$$
$$S={\p\over2}{\sqrt{ab}\,(a+c)(b+c)\over a+b}.\eqno(3.7)$$
They can be written in terms of physical quantities using the relations
$$a=2(M-D),\qquad b=2(M+D),\qquad c=2\sqrt{M^2+3D^2-Q^2-P^2}.\eqno(3.8)$$
A detailed discussion of the thermodynamics  is given in ref.~[26].
\bigbreak
\section{4. Special case $\f=0$}
There are no \bh solutions of our model with vanishing $\f$. However, it is possible to consider the limit where $\f$ decouples, as is often
assumed in Kaluza-Klein theories. Although this is not consistent with the field equations, it can be achieved if one adds a Lagrange multiplier
to the action [29].

Solutions of the model without scalar fields have been studied in [28] and in a simplified version, exact solutions have been found in [27].
Their properties depend on the sign of $\a$. Unfortunately, in those papers some signs were wrong; in the following we report the
correct results.\footnote{$^2$} {Anyway,
the results obtained in those paper are qualitatively correct, if one flips the sign of $\a$. In particular, solutions with everywhere regular electric
field exist for some values of the charges if $\a<0$.}

The solutions were calculated in a gauge different from the one used in sect.~3, namely $R=r$. In this gauge, the metric reads
$$ds^2=-\D(R)dt^2+{dR^2\over\s^2(R)\D(R)}+R^2d\O^2,\eqno(4.1)$$
with $\f=\f(R)$, and in terms of the \coo of (3.1), $\s={dR\over dr}$.

The \fe in this limit read [28]
$$\Bigg[A_0'R^2\s\Bigg(1-3\a{P^2\over R^4}-\a{\s^2\D-1\over R^2}\Bigg)\Bigg]'=0,$$
$$\D'R+\D+A_0'^2R^2-\s^{-2}\(1-{P^2\over R^2}\)+\a A_0'^2\bigg(1-3{P^2\over R^2}-3\s^2\D\bigg)+\a {P^2\over R^3}\D'=0,$$
$$\s'\(1+\a{P^2\over R^4}\)+{\a\over R}\bigg({3P^2\over R^4}+A'^2_0\s^2\bigg)=0,\eqno(4.2)$$
where a prime denotes derivative with respect to $R$. No analytic form is known for the solutions, but they were calculated in a perturbative
setting and numerically in [28].

It is well known that in the limit $\a=0$, the previous equations are solved by the \RN metric,
$$\D=\D_0\id1-{2M\over R}+{Q^2+P^2\over R^2},\qquad \s=1,\qquad A_0={Q\over R}.\eqno(4.3)$$

For $\a\ne 0$, the field equations admit a first integral, that gives an exact solution for  $A_0$,
$$E=A_0'={QR^2\over\s[R^4-\a\(\s^2\D-1\)R^2-3\a P^2]},\eqno(4.4)$$
where $Q$ is an integration constant that can be identified with the electric charge.

However, the other equations cannot be solved exactly. One can use a perturbative expansion in
the small parameter $\a$ around the \RN background.
The expansion will be valid for large values of $R$, namely $R\gg\sqrt\a$.


One can define the perturbations $\s(R)$, $\g(R)$ and $\f(R)$ through an expansion at order $\a$ around the RN background,
$$\D=\D_0+\a\d,\qquad\s=1+\a\g,\qquad E={Q\over R^2}(1+\a\q).\eqno(4.5)$$
Substituting in the \fe and choosing \bc such that the corrections vanish at infinity, one obtains up to first order in $\a$ the solutions
$$\D\sim1-{2M\over R}+{Q^2+P^2\over R^2}+\a\({P^2-Q^2\over2R^4}+{M(P^2+Q^2)\over2R^5}-{7P^4+4P^2Q^2+3Q^4\over10R^6}\),\eqno(4.6)$$

$$\s\sim1+\a\,{Q^2+3P^2\over4R^4},\eqno(4.7)$$

$$E\sim{Q\over4R^2}\(1-{8\a M\over R^3}+{\a(3Q^2+13P^2)\over R^4}\).\eqno(4.8)$$
Our approximation is good for $R\to\inf$. At leading orders in $1/R$ the \ab is the same as in RN. Therefore, we can still identify
$M$ with the mass of the black hole and $Q$ and $P$ with its electric and magnetic charge.

In this approximation, the horizons are displaced with respect to the RN horizons $r_\pm$ by a quantity $\a\e_\pm$, with
$$\e_\pm=-{\d\over\D_0'}\Bigg|_{R=R_\pm}={5(7Q^2-27P^2)r_\pm^2+3(13Q^4-44Q^2P^2+27P^4)\over60r_\pm^3(r_\pm^2-Q^2-P^2)}$$
Thermodynamical quantities can also be computed, see [28] for details.

In [28] the solutions were calculated also numerically. For comparison with the results of next section, in figures 1-4 we plot the graphs of $\D$
and $\s$ for selected values of $Q$ and $P$, both for positive and negative values of the coupling constant.
Clearly, all the solutions have identical behaviour for large values of $R$, coinciding with that of the RN metric,
while they greatly differ from it only very close to the singularity. All of them present two horizons that shield a curvature singularity.
However, while for positive $\a$, the metric functions have finite values at the singularity, for negative $\a$ both $\D$ and $\s$ diverge
for some values of the charges.

It must be noted however that the behaviour of the solutions depends strongly on the value of the coupling constant and of the charges, and is
more varied than it may appear from the few examples proposed.

\bigskip
\centerline{\epsfysize=4.5truecm\epsfbox{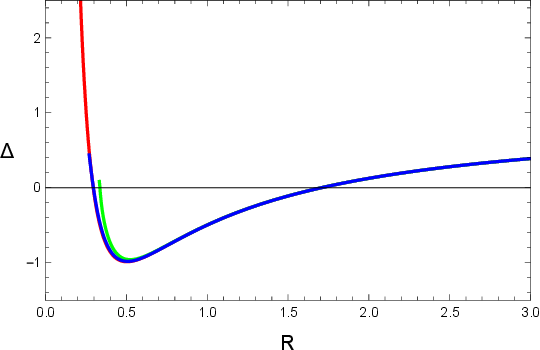}\qquad \epsfysize=4.5truecm\epsfbox{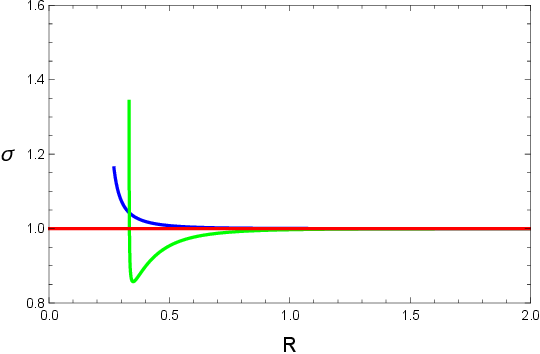}}
\smallskip
\baselineskip10pt{\noindent{\smalll Fig.\ 1: The metric function $\scriptstyle{\D}$ and  $\scriptstyle{\s}$ for black holes with $\scriptstyle{M\ug1}$ and $\scriptstyle{Q\ug0.66}$, $\scriptstyle{P\ug0.26}$
(blue), $\scriptstyle{Q\ug0.26}$, $\scriptstyle{P\ug0.66}$  (green) with $\scriptstyle{\a\ug0.01}$, compared with the RN solution (red).  In the first case, a singularity is
located at $\scriptstyle{R\ug0.27}$ and the horizons at $\scriptstyle{R\ug0.30}$ and $\scriptstyle{R\ug1.71}$. In the second case the singularity occurs at $\scriptstyle{R\ug0.33}$ and the horizons at
$\scriptstyle{R\ug0.34}$ and $\scriptstyle{R\ug1.71}$. Finally,
for RN the singularity is at the origin and two  horizons occur  at $\scriptstyle{R\ug0.30}$ and $\scriptstyle{R\ug1.71}$.}

\bigskip
\centerline{\epsfysize=4.5truecm\epsfbox{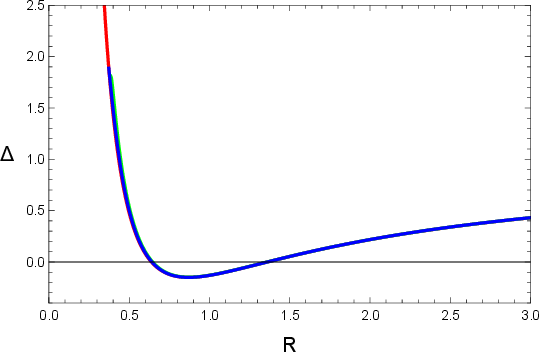}\qquad \epsfysize=4.5truecm\epsfbox{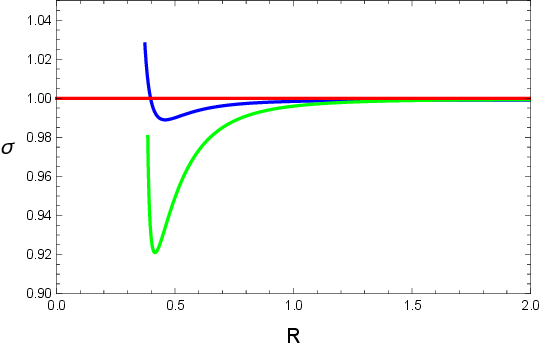}}
\smallskip
\baselineskip10pt{\noindent{\smalll Fig.\ 2: The metric function $\scriptstyle{\D}$ and  $\scriptstyle{\s}$ for black holes with $\scriptstyle{M\ug1}$ and $\scriptstyle{Q\ug0.76}$, $\scriptstyle{P\ug0.54}$
(blue), $\scriptstyle{Q\ug0.54}$, $\scriptstyle{P\ug0.76}$  (green) with $\scriptstyle{\a\ug0.01}$, compared with the RN solution (red). In the first case, the horizons are
located at $\scriptstyle{R\ug0.63}$ and $\scriptstyle{R\ug1.37}$, and a singularity at $\scriptstyle{R\ug0.23}$. In the second case the horizons are placed at $\scriptstyle{R\ug0.63}$ and $\scriptstyle{R\ug1.37}$
and the singularity at $\scriptstyle{R\ug0.28}$. Finally, for RN the horizons lie at $\scriptstyle{R\ug0.64}$ and $\scriptstyle{R\ug1.37}$ and the singularity is at the origin.}
\bigskip

\centerline{\epsfysize=4.5truecm\epsfbox{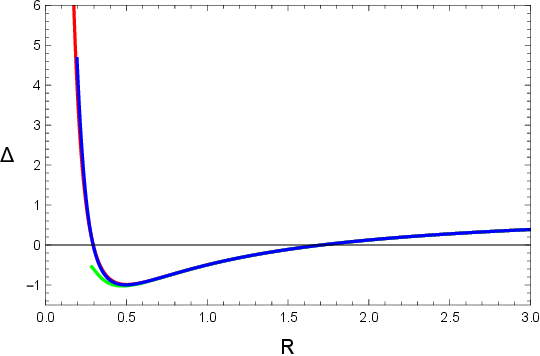}\qquad \epsfysize=4.5truecm\epsfbox{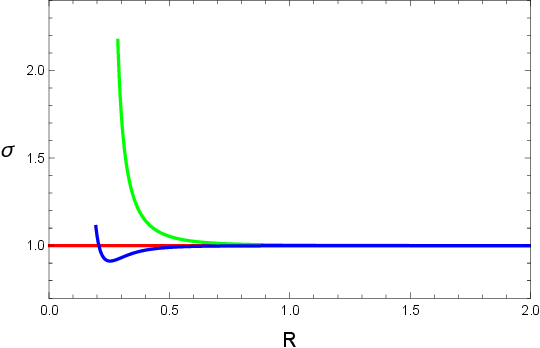}}
\smallskip
\baselineskip10pt{\noindent{\smalll Fig.\ 3: The metric function $\scriptstyle{\D}$ and  $\scriptstyle{\s}$ for black holes with $\scriptstyle{M\ug1}$ and $\scriptstyle{Q\ug0.66}$, $\scriptstyle{P\ug0.26}$
(blue), $\scriptstyle{Q\ug0.26}$, $\scriptstyle{P\ug0.66}$  (green) with $\scriptstyle{\a\ug-\,0.01}$, compared with the RN solution (red). In the first case, the horizons are
located at $\scriptstyle{R\ug0.29}$ and $\scriptstyle{R\ug1.70}$, and the singularity at $\scriptstyle{R\ug0.20}$. In the second case there is a unique horizon at $\scriptstyle{R\ug1.69}$, while
the singularity is at $\scriptstyle{R\ug0.29}$. Finally, for RN the singularity is at the origin and two  horizons occur  at $\scriptstyle{R\ug0.30}$ and $\scriptstyle{R\ug1.71}$.}
\bigskip
\bigbreak
\centerline{\epsfysize=4.5truecm\epsfbox{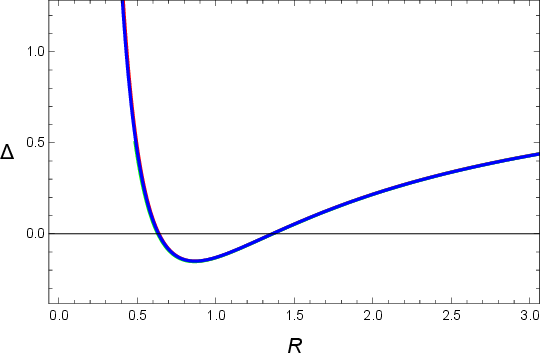}\qquad \epsfysize=4.5truecm\epsfbox{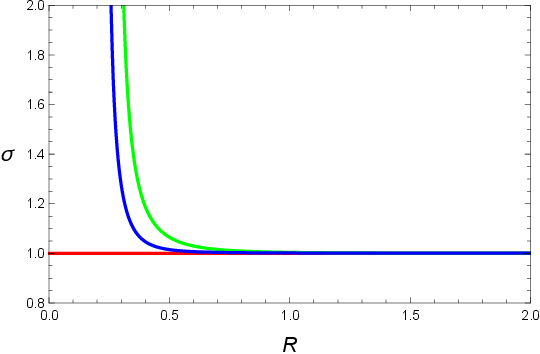}}
\smallskip
\baselineskip10pt{\noindent{\smalll Fig.\ 4: The metric function $\scriptstyle{\D}$ and  $\scriptstyle{\s}$ for black holes with $\scriptstyle{M\ug1}$ and $\scriptstyle{Q\ug0.76}$, $\scriptstyle{P\ug0.54}$
(blue), $\scriptstyle{Q\ug0.54}$, $\scriptstyle{P\ug0.76}$  (green) with $\scriptstyle{\a\ug-\,0.01}$, compared with the RN solution (red).In the first case, the horizons are
located at $\scriptstyle{R\ug0.63}$ and $\scriptstyle{R\ug1.37}$, and the singularity at $\scriptstyle{R\ug0.23}$. In the second case they occur at $\scriptstyle{R\ug0.63}$ and $\scriptstyle{R\ug1.37}$  and the
singularity at $\scriptstyle{R\ug0.28}$. Finally, for RN the singularity is at the origin and two  horizons occur  $\scriptstyle{R\ug0.64}$ and  $\scriptstyle{R\ug1.37}$.}
\bigskip
\baselineskip12pt

It can also be checked that when $\a<0$ the electric field is regular at the curvature singularity for a range of values of the charges.

\section{5. General case}
We now consider solutions with nontrivial scalar field and $\a\ne0$. In this case, even a perturbative calculation is not feasible,
so one has necessarily to resort to numerical calculations.

The field equations can be obtained by varying the action with respect to the fields, and then fixing the gauge.
One obtains four independent equations, that have a very awkward form. They are reported in the appendix.
From those results it is evident that now the electric field has a much more involved behaviour than (4.4).
The system of differential equations reported in the Appendix can be solved numerically, starting from \bc such that near the horizon the solutions are close to the exact $\a=0$
solutions (3.2), (3.3), and then modifying them in order to obtain the correct asymptotic behaviour $\D\to1$, $\s\to1$, $\f\to0$ for $R\to\infty$.

It turns out that the solutions possess at most two horizons, that shield a singularity.
In figures 5-8 the plots of the metric functions and of the scalar field with $\a=\pm\,0.01$ for selected values of the electric and magnetic charge (equal to those of figs.~1-4)
are compared with the DM solutions with vanishing $\a$. Unfortunately, our numerical evaluation breaks down at the inner horizon, so we plot only the region exterior to it.
Moreover, since asymptotically all solutions tend to 1 and have identical behaviour, we plot their graphs only in the region near the horizon, where they differ sensibly.

\bigbreak
\centerline{\epsfysize=4.5truecm\epsfbox{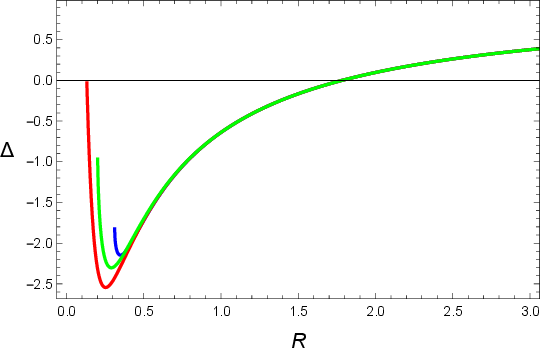}\qquad \epsfysize=4.5truecm\epsfbox{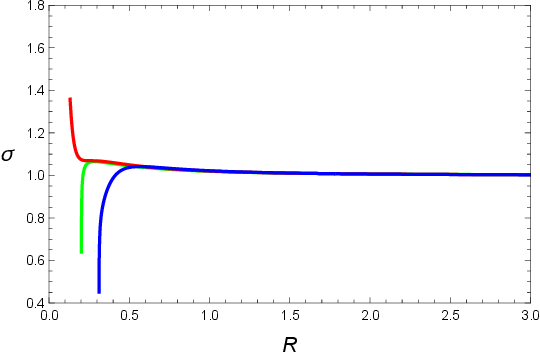}}
\bigskip
\centerline{\epsfysize=4.5truecm\epsfbox{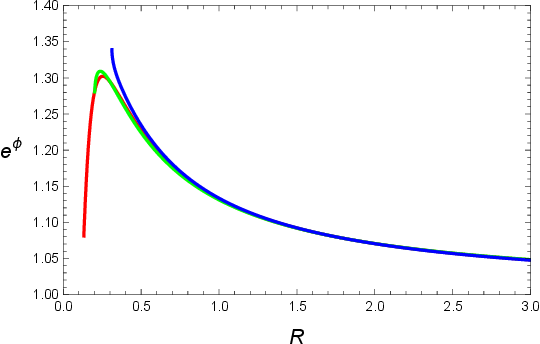}}
\smallskip
\baselineskip10pt{\noindent{\smalll Fig.~5: The metric function (a) $\scriptstyle{\D(R)}$, (b) $\scriptstyle{\s(R)}$, for
$\scriptstyle{M\ug1}$, with $\scriptstyle{\a\ug0.01}$: in blue $\scriptstyle{Q\ug0.66}$, $\scriptstyle{P\ug0.26}$, in green $\scriptstyle{Q\ug0.26}$, $\scriptstyle{P\ug0.66}$,
while for comparison we plot in red $\scriptstyle{\a\ug0}$ (in this case the solutions coincide because of the duality, except that $\scriptstyle{\f\;\to\;-\f}$).
In (c)  $\scriptstyle{e^\f}$ in the first case, $\scriptstyle{e^{-\f}}$ in the second case, together with $\scriptstyle{e^\f}$ for $\scriptstyle{\a\ug0}$. The outer horizon is fixed at $\scriptstyle{R\ug1.73}$.
If $\scriptstyle{\a\;\ne\;0}$ the metric function $\scriptstyle{\D(r)}$ presents a singularity at $\scriptstyle{R\ug0.33}$ for
$\scriptstyle{Q\ug0.66}$, $\scriptstyle{P\ug0.26}$ or at $\scriptstyle{R\ug0.27}$ for $\scriptstyle{Q\ug0.26}$, $\scriptstyle{P\ug0.66}$, while for $\scriptstyle{\a\ug0}$ a horizon occurs at
$\scriptstyle{R\ug0.20}$.}
\bigskip\bigskip
\centerline{\epsfysize=4.5truecm\epsfbox{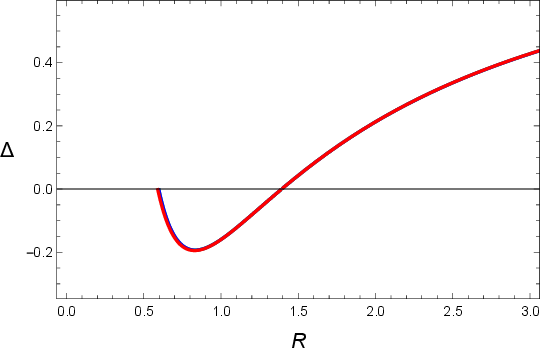}\qquad \epsfysize=4.5truecm\epsfbox{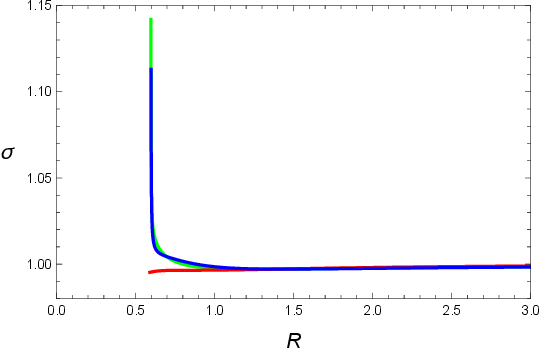}}
\bigskip
\centerline{\epsfysize=4.5truecm\epsfbox{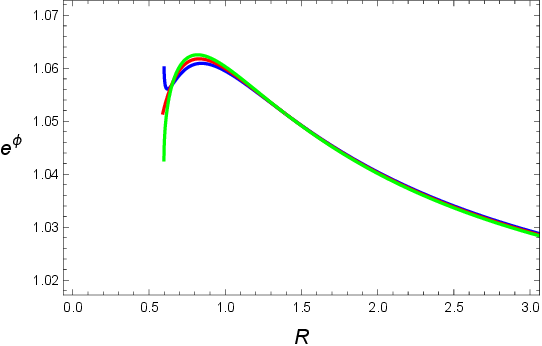}}
\smallskip
\baselineskip10pt{\noindent{\smalll Fig.~6:  The metric function (a) $\scriptstyle{\D(R)}$, (b) $\scriptstyle{\s(R)}$ for $\scriptstyle{M\ug1}$, with $\scriptstyle{\a\ug0.01}$: in blue $\scriptstyle{Q\ug0.76}$, $\scriptstyle{P\ug0.54}$, in green $\scriptstyle{Q\ug0.54}$, $\scriptstyle{P\ug0.76}$,
while in red $\scriptstyle{\a\ug0}$ (in this case the solutions coincide because of the duality, except that $\scriptstyle{\f\;\to\;-\f}$.
In (c)  $\scriptstyle{e^\f}$ in the first case, $\scriptstyle{e^{-\f}}$ in the second case, together with $\scriptstyle{e^\f}$ for $\scriptstyle{\a\ug0}$.  The horizon is fixed at $\scriptstyle{R\ug1.40}$.
In this case, the metric functions $\scriptstyle{\D(R)}$ are almost identical in all cases, and the inner horizon is located at $\scriptstyle{R\ug0.60}$; for $\scriptstyle{\a\ug0}$ the singularity is at
$\scriptstyle{R\ug0.12}$.}

\bigbreak
\centerline{\epsfysize=4.5truecm\epsfbox{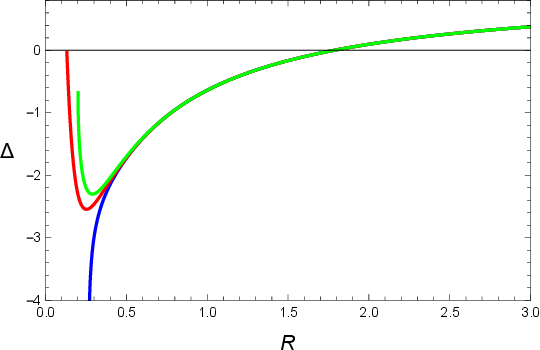}\qquad \epsfysize=4.5truecm\epsfbox{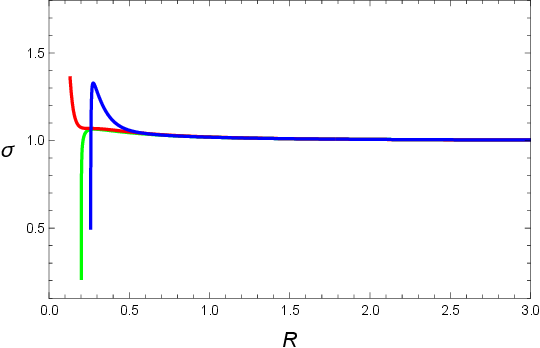}}
\bigskip
\centerline{\epsfysize=4.5truecm\epsfbox{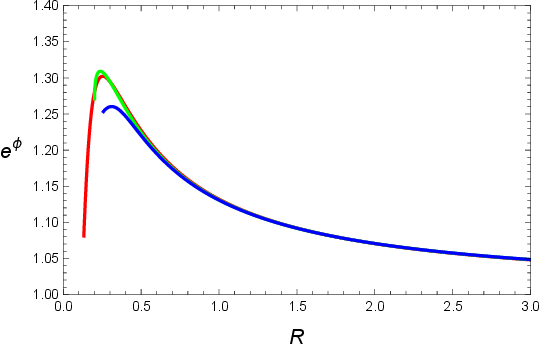}}
\smallskip
\baselineskip10pt{\noindent{\smalll Fig.~7: As fig.~5, but with $\scriptstyle{\a\ug-0.01}$.
The horizon is still fixed at $\scriptstyle{R\ug1.73}$ and the singularities are at $\scriptstyle{R\ug0.26}$
for $\scriptstyle{Q\ug0.66}$, $\scriptstyle{P\ug0.26}$ and at $\scriptstyle{R\ug0.20}$  for $\scriptstyle{Q\ug0.26}$, $\scriptstyle{P\ug0.66}$.}
\bigskip\bigskip
\centerline{\epsfysize=4.5truecm\epsfbox{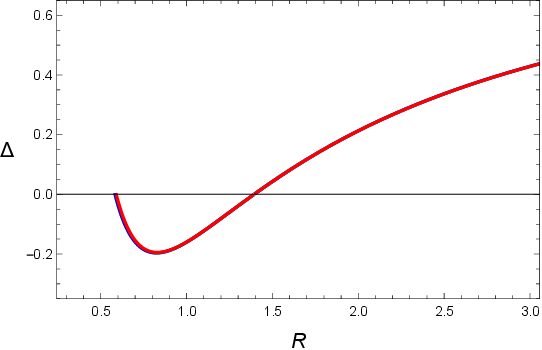}\qquad \epsfysize=4.5truecm\epsfbox{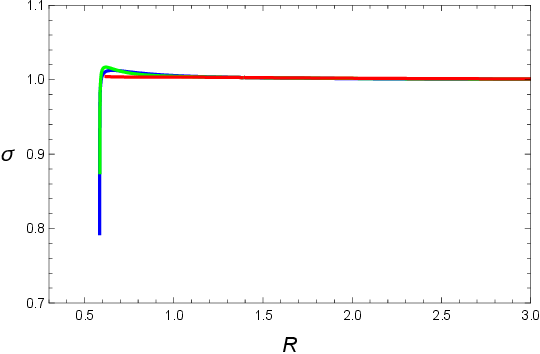}}
\bigskip
\centerline{\epsfysize=4.5truecm\epsfbox{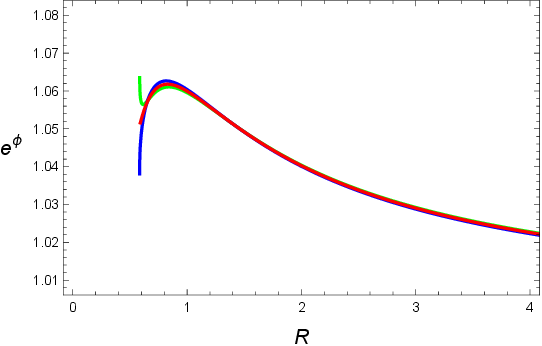}}
\smallskip
\baselineskip10pt{\noindent{\smalll Fig.~8: As fig.~6, but with $\scriptstyle{\a\ug-0.01}$. The horizon is still fixed at $\scriptstyle{R\ug1.40}$.
Also in this case the metric functions $\scriptstyle{\D(R)}$ are almost indistinguishable; the inner horizon is located at $\scriptstyle{R\ug0.59}$.}

\bigskip
\baselineskip12pt

From our results it appears that for a range of values of $Q$ and $P$ (figs. 6 and 8), the behaviour of the solutions is analogous to that of the DM solutions, with two
 horizons, while in other regions of the parameter space (figs. 5 and 7), the singularity arises  before the inner horizon is reached.
 In the first case, the function $\D$ is almost indistinguishable from the $\a=0$ solution.
The behaviour of the metric function $\s$ and of the scalar field is more varied, but both take finite values at the inner horizon.
It is also evident that the form of the solutions does not depend much on the sign of $\a$, except for the behaviour of $\s$ at small values of $R$.
Comparing with the solutions of sect.~4, where the scalar field has been neglected, we see that although they present qualitative similarities, several features
are different.

As in section 4, the figures do not exhaust all the possible forms of the solutions, since these depend on three parameters and therefore it is not possible to show all
the different possibilities, but are representative of their generic behaviour.

As in the DM case, a minimal allowed mass $M_{ext}$, corresponding to extremal black holes, is present for given values of the charges, but in our case  $M_{ext}^2$ can
be larger than $Q^2+P^2$, contrary to the DM solutions.
Moreover, as for ordinary 4-dimensional dilatonic Einstein-GB black holes [18-21], a minimal mass is present also if $Q=P=0$.

\bigbreak
\section{6. Thermodynamics}
The behaviour of the thermodynamical quantities can be determined numerically starting from the solutions of sect.~5.

The temperature can be calculated as the inverse periodicity of the Euclidean section  by the standard formula
$$T={1\over4\p}{g'_{tt}\over\sqrt{g_{tt}g_{RR}}}\Bigg|_{R=R_h}={\s\D'\over4\p}\Big|_{R=R_h},\eqno(6.1)$$
with $R_h$ the location of the outer horizon.

The Wald entropy is defined as [30]
$$S=-{1\over8}\int_\S{\de\cL\over\cR_{\m\n\r\s}}\e_\mn\e_{\r\s} d\O^2,\eqno(6.2)$$
where $\e_\mn$ is the binormal vector to the bifurcation surface $\S$ and the integration is performed over the horizon.
In our case,
$${\de\cL\over\cR_{\m\n\r\s}}=g^{\m\r}g^{\n\s}+2\a\[\cR^{\m\n\r\s}-4g^{\m\r}\cR^{\n\s}+g^{\m\r}g^{\n\s}\cR)e^{-2\f}-(F^\mn F^{\r\s}
-4g^{\m\r}F^{\n\l}F^\s_{\ \l}+g^{\m\r}g^{\n\s}F^2)e^{-8\f}\].\eqno(6.3)$$
\bigskip
It follows that
$${\de\cL\over\cR_{\m\n\r\s}}\,\e_\mn\e_{\r\s}=-2-8\a\({1-\D\s^2\over R^2}\,e^{-2\f}-{P^2\over R^4}\,e^{-8\f}\).\eqno(6.4)$$
Substituting in (6.2), one gets
$$S={A\over4}\[1+4\a\({1-\D\s^2\over R^2}\,e^{-2\f}-{P^2\over R^4}\,e^{-8\f}\)\bigg|_{R=R_h}\ \]=
{A\over4}\[1+4\a\({e^{-2\f_h}\over R_h^2}-{P^2e^{-8\f_h}\over R_h^4}\) \],\eqno(6.5)$$
$\f_h$ being the value of the dilaton at the horizon.

In fig.~9, the dependence of the temperature and entropy on the mass for positive $\a$ is depicted for several values of the charges.
The curves end at the minimal mass.
While the behaviour of the entropy is essentially the same for any value of the charges, that of the temperature depends more
decidedly on $Q$ and $P$.  The temperature has a maximum for given value of $M$,
except in the cases in which either $Q$ or $P$ vanish, when this is reached at the extremal mass.
Asymptotically, the behaviour of the temperature becomes almost indistinguishable for different charges,  while for small
$M$ the graphs bifurcate when one exchanges the values of $Q$ and $P$. For negative $\a$ the plots are very similar, so we do not show them.

Compared with the DM case, the main difference is that the temperature of extremal \bh does not vanish, but takes a finite value,
similarly to the pure dilatonic GB model [20].

\bigskip
\centerline{\epsfysize=4.5truecm\epsfbox{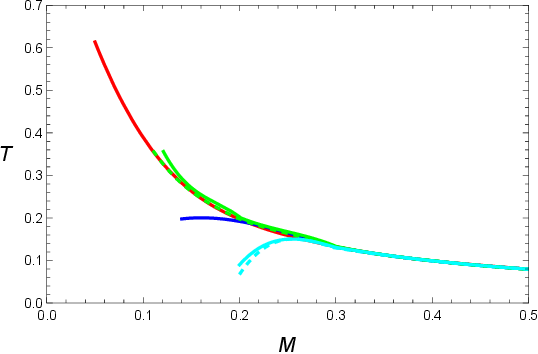}\qquad \epsfysize=4.5truecm\epsfbox{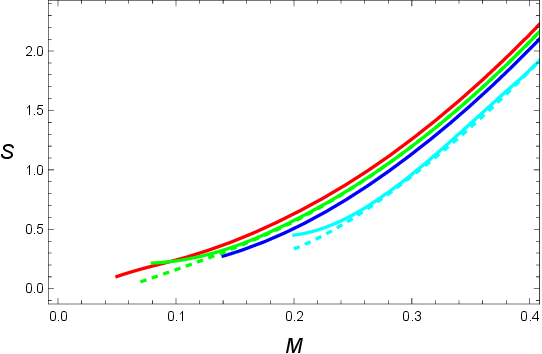}}
\medskip
\baselineskip10pt{\noindent{\smalll Fig.~9:  The figures display the values of the temperature and of the entropy as a function of
the mass for selected values of the charges: $\scriptstyle{Q\ug P\ug0}$ (red), $\scriptstyle{Q\ug P\ug0.1}$ (blue),
$\scriptstyle{Q\ug0.1,\ P\ug0}$ (green), $\scriptstyle{Q\ug0.2,\ P\ug0.1}$ (cyan). Dashed curves correspond to the exchange of the previous
values of $\scriptstyle{Q}$ and $\scriptstyle{P}$.
The curves end at the minimal values of the mass.
Exchanging $\scriptstyle{Q}$ and $\scriptstyle{P}$ modifies the thermodynamic functions only for values of the mass near extremality.
For $\scriptstyle{M\;>\;0.3}$ the temperature is essentially the same for all values of the charges.
Notice that the case $\scriptstyle{Q\ug P\ug0}$ does not entail a minimal mass.}
\bigbreak

\baselineskip12pt

\section{7. Conclusions}
We have investigated black holes solutions of dimensionally reduced 5D gravity with Einstein-GB action,
generalizing the results of [28], where the scalar field had been neglected. In absence of the GB term, the solutions
reduce to those found long time ago by Dobiasch and Maison [23].
Our solutions differ sensibly from those only in the interior region of the hole.

The electric-magnetic duality is no longer satisfied, because of the presence of the GB term, but the differences between solutions with interchanged
charges are not very large.
Also the other properties of the solutions are similar to DM, but depend more strongly on the values of the charges and possess only one horizon
for a range of values of $Q$ and $P$.

For what concerns thermodynamics,
the holes become extremal for values of the mass that can be larger than in the DM case, and their temperature does not vanish. A minimal mass occurs
also for vanishing charges, as in similar models [18-21].
The behaviour of the entropy is more similar to that occurring in DM. Again exchanging $Q$ and $P$ the thermodynamical functions are only slightly
modified.

\section{Appendix}
In the gauge (4.1) the equations for the electric potential and the metric
components can be written as

$$\Bigg[A_0'\s\Bigg(R^2e^{-6\f}+\a\(e^{-8\f}-3{P^2\over R^2}e^{-14\f}\)-\a\s^2\D\(1+R^2\f'^2\)e^{-8\f}\Bigg)\Bigg]'=0,\eqno(A.1)$$
$$\eqalignno{&\D'R+\D-3\D R^2\f'^2+A_0'^2R^2e^{-6\f}-\s^{-2}\(1-{P^2\over R^2}e^{-6\f}\)+\a\bigg[-3{P^2A_0'^2\over R^2}e^{-14\f}
-(\D'\f'\cr&+2\D\f'^2)e^{-2\f}+A_0'^2e^{-8\f}+{P^2\over R^3}\big(\D'+4\D'R\f'+8\D R\f'^2+2\D\f'\big)e^{-8\f}\bigg]
+3\a\s^2\D\bigg[\big(\D'\f'\cr&+2\D'R\f'^2+2\D\f'^2+\D'R^2\f'^3+4\D R\f'^3+2\D R^2\f'^4\big)e^{-2\f}-A_0'^2(1+R^2\f'^2)e^{-8\f}\bigg]=0,&(A.2)}$$
$$\eqalignno{&\s'R +3R^2\s\f'^2+\a\s'\[-\f'e^{-2\f}+{P^2\over R^3}\(1+4R\f'\)e^{-8\f}\]+\a\s\bigg[\(-\f''+4\f'^2\)e^{-2\f}+{P^2\over R^4}\Big(4R^2\f''\cr&
-40R^2\f'^2-18R\f'-3\Big)e^{-8\f}\bigg]+\a\s^2\bigg[\s\D(1+4R\f'+3R^2\f'^2)\f''-2\s\D(2+5R\f'+3R^2\f'^2)\f'^2\cr&+3\s'\D(\f'+2R\f'^2+R^2\f'^3)\bigg]e^{-2\f}
+\a\s^3 A_0'^2(1+R^2\f'^2)e^{-8\f}=0,&(A.3)}$$

$$\eqalignno{&2\s\D'+R\s'\D'+2\s'\D+R\s\D''+6R\s'\D\f'-2\({P^2\over\s R^3}+A_0'^2R\s\)e^{-6\f}+\a{P^2\over R^3}\s'(\D'+2\D\f')e^{-8\f}\cr
&+\a{P^2\over R^3}\s\Big[-6A_0'^2e^{-14\f}+(\D''+2\D'\f'+2\D\f'')e^{-8\f}\Big]+6\a\s^2\s'\D\big(R\D'\f'^2+\D\D'\f'+2\D^2\f'^2\cr&
+2R\D+A_0'^2\D\big)e^{-2\f}+\a\s^3\Big[\big(2R\D'^2\f'^2+2R\D\D'\f'^3-12R\D^2\f'^4+2\D'^2\f'+4\D\D'\f'^2+8\D^2\f'^3\cr&
+2R\D\D''\f'^2+2\D\D''\f'+4R\D\D'\f''+8\D^2\f'\f''+12\D^2\f'\f''+2\d\d'\f'^4\big)e^{-2\f}+A_0'\Big(2A_0'(R\D\f'^2\cr&
-\D'+8\D\f')+4A_0''\D\Big)e^{-8\f}\Big]=0.&(A.4)}$$
Finally, we write the scalar equation in the form
$$\y_0\f''+\y_1\f'+\y_2\f'^2+\y_3\f'^3+\y_4=0,\eqno(A.5)$$
where
$$\eqalign{\y_0&=\s\D\Bigg[-6R^2+\a\bigg[-4e^{-2\f}+\({16P^2\over R^2}-2\s^2A_0'^2R^2\)e^{-8\f}+\s^2\Big(4\D+4R\D'+24R\D\f'\cr&+6R^2\D'\f'+24R^2\D\f'^2\Big)e^{-2\f}\bigg]\Bigg],\cr
\y_1&=-6(2\s R\D+\s R^2\D'+\s'R^2\D)+\a\bigg[{16P^2\over R^3}(\s R\D'-2\s\D+\s'R\D)e^{-8\f}-2A_0'\s^3R(2A_0'\D\cr&+A_0'R\D'+2A_0''R^2\D)e^{-8\f}-6A_0'^2\s^2\s'R^2\D e^{-8\f}-4(\s\D'+\s'\D)e^{-2\f}+\s^3(12\D\D'\cr&+4R\D'^2+4R\D\D'')e^{-2\f}+3\s^2\s'(4\D^2+4R\D\D')e^{-2\f}\bigg],\cr
\y_2&=\a\s\bigg[4\D e^{-2\f}-8\D\(8{P^2\over R^2}-A_0'^2\s^2R^2\)e^{-8\f}+\s^2(3R^2\D\D''+3R^2\D'^2+26R\D\D'\cr&+8\D^2)e^{-2\f}+9\s\s'R(4\D^2+R\D\D')e^{-2\f}\bigg],\cr
\y_3&=4\a\s^2R\D(4\s\D+3\s R\D'+6\s'R\D)e^{-2\f},\cr
\y_4&=-12\a\s^3R^2\D^2e^{-2\f},\cr
\y_5&=-6\({P^2\over\s R^2}-\s A_0'^2R^2\)e^{-6\f}+\a\bigg[\Big((-\s\D''+\s^3(\D\D''+\D'^2)+\s'(-\D'+3\s^2\D\D')\Big)e^{-2\f}\cr&-\Big(
8A_0'^2\s(1+\s^2\D)+2\s{P^2\over R^4}\(3\D-R\D'+2R^2\D''\)+2\s'{P^2\over R^3}(\D+2R\D')\Big)e^{-8\f}\cr&-42\s{A_0'^2P^2\over R^2}e^{-14\f}\bigg].}$$
Eq.~(A.1) can be immediately integrated to get
$$A_0'=E={QR^2e^{6\f}\over\s\[R^4+\a\(R^2e^{-2\f}-3P^2e^{-8\f}\)-\a\s^2\D R^2\(1+R^2\f'^2\)e^{-2\f}\]},\eqno(A.6)$$
where $Q$ is the electric charge. One can then substitute the result for $A_0'$ in the previous equations.

Only three of the equations (A.2-A.5) are independent. Combining (A.4) and (A.5) one can eliminate $\D''$ from the system, obtaining a system of three differential
equations with the highest order derivatives given by $\f''$, $\D'$ and $\s'$, that can be solved numerically imposing suitable boundary conditions.

\bigskip

\section{Acknowledgments}
 We wish to thank Andrea P. Sanna for pointing out an error in a previous version of the manuscript and for useful comments.
S. Mignemi acknowledges support from GNFM.
\beginref
\ref [1] T. Kaluza, Sitz. Preuss. Akad. Wiss., Math. Phys. {\bf 1}, 966 (1921).
\ref [2] O. Klein, Z. Phys.\ {\bf 37}, 895 (1926).
\ref [3] P. Jordan, Naturwissenschaften {\bf 11}, 250 (1946).
 M.Y. Thiry, Compt. Rend. Acad. Sci. Paris {\bf 226}, 216 (1948).
\ref [4] J. Scherk and J. Schwarz, \NP{B153}, 61 (1979).
\ref [5] D. Lovelock, \JMP{12}, 498 (1971).
\ref [6] H.A. Buchdal, \JoP{A12}, 1037 (1979).
\ref [7]  R. Kerner, C.\ R.\ Acad.\ Sc.\ Paris {\bf 304}, 621 (1987).
\ref [8] J. Pleba\'nski, {\it Lectures on non-linear electrodynamics}, NORDITA, Copenhagen, 1968.
\ref [9] F. M\"uller-Hoissen, \NP{B337}, 709 (1990).
\ref [10] G.W. Horndeski, \IJTP{10}, 363 (1974).
G.W. Horndeski and J. Wainwright, \PR{D16}, 1691 (1977).
\ref [11] B. Zwiebach, \PL{B156}, 315 (1985). B. Zumino, \PRep{137}, 109 (1986).
\ref [12] J.D. Bekenstein, \PR{D5}, 1239; 2403 (1972).
 W. Israel, \PR{164}, 1776 (1967).
\ref [13] H. Luckock and I.G. Moss, \PL{B176}, 341 (1986).
\ref [14] M.S. Volkov and D.V. Gal'tsov,  JETP Lett.\ {\bf 50} 346 (1989).
\ref [15] G. Gibbons and K. Maeda, \NP{B298}, 741 (1988).
\ref [16] D. Garfinkle, G.T. Horowitz and A. Strominger, \PR{D43}, 3140 (1991).
\ref [17] C.F.E. Holzhey and F. Wilczek, \NP{B380}, 447 (1992).
\ref [18] S. Mignemi and N.R. Stewart, \PR{D47}, 5259 (1993).
\ref [19] P. Kanti, N.E. Mavromatos, J. Rizos, K. Tamvakis and E. Winstanley, \PR{D54}, 5049 (1996).
\ref [20] T.  Torii,  H.  Yajima, and K. Maeda, \PR{D55}, 739 (1997).
\ref [21] S.O. Alexeyev and M.V. Pomazanov, \PR{D55}, 2110 (1997).
\ref [22] S. Coleman, J. Preskill and F. Wilczek, \PRL{67}, 1975 (1991);
\ref [23] P. Dobiasch and D. Maison, \GRG{14}, 231 (1982).
\ref [24] A. Chodos  and S. Detweiler, \GRG{14}, 870 (1982).
\ref [25] D. Pollard, \JoP{A16}, 565 (1983).
\ref [26] G.W. Gibbons and D. Wiltshire, \AoP{167}, 201 (1986).
\ref [27] S. Mignemi,  \IJMP{A37}, 2250065 (2022).
\ref [28] S. Mignemi,  Universe {\bf 9}, 509 (2023).
\ref [29]  H.H. Soleng and O. Gron, \AoP{240}, 432 (1995).
\ref [30] R.M. Wald, \PR{D48}, R3427 (1993).

\bigskip

\endref
\end

It is also interesting to investigate the behavior of the solutions near the singularity. This can be calculated by an expansion
in powers of $r$ near $r=0$. Setting
$$\D\sim R^h,\qquad \s\sim R^l,\qquad E\sim R^k,$$
and substituting in the \fe (13)-(15), one obtains $h=-1$, $l=-3$ and $k=-1$.
It follows that the metric functions and the electric field diverge for $r=0$.

The only exception is for $P=0$. In this case, the electric field vanishes at the origin, like in the GHS solution.
In fact now $h=-1$, $l=0$ and $k=1$. The existence of electric solutions regular at the origin in absence of magnetic field had  also
been noticed in [sg]. However, for small values of $Q$, numerical computations show that the solutions become singular at a point $r_0>0$,
presenting a spherical singularity, similarly with the $\a<0$ solutions of sect.~2.

The horizons are displaced with respect to the RN solutions, where they are located at $r^\ast_\pm$.
At first order in $\a$, one has $r_\pm=r^\ast_\pm+\a\D r_\pm$, where
$$\D r_\pm=-{\s\over A'}\Big|_{r=r^\ast_\pm}.\eqno(28)$$
It follows that
$$\D r_\pm={5(P^2-Q^2)\rpm^2-3 P^4 + 8 P^2 Q^2 - Q^4\over5\rpm^3 (\rpm^2-P^2 - Q^2)},\eqno(29)$$
where to simplify the expression we have chosen as independent parameters $r^\ast_+\,(r^\ast_-)$, $Q$ and $P$.
The actual values of these displacements strongly depend on the  charges.

A calculation shows that the condition of extremality is, at first order in $\a$,
$$M^2=P^2+Q^2+{\a\over5}\,{P^4+4Q^2P^2-3Q^4\over (P^2+Q^2)^2}.\eqno(30)$$
Depending on the values of $Q$ and $P$ the correction with respect to the RN case can be both positive or negative.
We recall however that, while the value of $r_+$ obtained in this way is in general well approximated by (29), the value of $r_-$  is reliable
only for very small values of $\a$.

In this approximation, the metric function $e^{2\n}$ does not differ much from that of RN and the causal structure should therefore be analog.
Hence, for $M$ greater than extremality, one has two horizon, while a naked singularity is present for $M$ less than its extremal value.
However, this is not necessarily true for greater values of $\a$, where the approximation fails.
\medskip

Using standard definitions it is possible to derive the thermodynamical quantities associated to the black hole
from the behaviour of the metric functions near the outer horizon.
The temperature can be calculated from the formula
$$T={1\over4\p}\,e^{-\la}(e^{2\n})'\big|_{r=r_+}\sim {1\over4\p}\Big[A'(1-\a\g)+\a\s'\Big]_{r=r_+}.\eqno(31)$$
Hence,
$$\eqalign{ T\sim&\ {1\over4\p\rp^3}\bigg(\rp^2-P^2-Q^2\cr
&-2\a\ {7 P^6 + 25 P^4 Q^2 + 17 P^2 Q^4 - Q^6 - 14 P^4\rp^2 -
36 P^2 Q^2\rp^2 + 2 Q^4\rp^2 + 5(Q^2+P^2)\rp^4\over 5\rp^4(\rp^2-P^2-Q^2)}\bigg).}\eqno(32)$$

The entropy $S$ is usually identified with the area of the horizon, namely,
$$S\sim4\p\rp^2\(1-2\a{3P^4-8P^2Q^2+Q^4+5(Q^2-P^2)\rp^2\over5\rp^4(\rp^2-P^2-Q^2)}\).\eqno(33)$$
It follows that the thermodynamical quantities display a complicate dependence on the charges.

\ref [1] R. Kerner, Ann. Inst. H. Poincar\'e Phys.Theor. 9, 143 (1968).
\ref [2] R. Kerner, C.\ R.\ Acad.\ Sc.\ Paris {\bf 304}, 621 (1987).

\ref [6] R. Kerner, in {\it Proceedings of Varna Summer School, "Infinite Dimensional Lie Algebras and Quantum Field Theory"},
World Scientific 1987, \arx{2303.10603}.

\ref [13] F. M\"uller-Hoissen, \PL{B201}, 325 (1998).
A. Shapere, S. Trivedi and F. Wilczek,
\MPL{A6}, 2677 (1991).
horowitz wiseman
R. Metsaev and A. Tseytlin, Nucl. Phys. B293, 385 ~1987!.
D.G. Boulware, S. Deser, \PL{B175}, 409 (1986).
D.L. Wiltshire, \PR{D38}, 2445 (1988).

$$\eqalignno{&\s\D\Bigg[-6r^2+\a\bigg[-4e^{-2\f}+\({16P^2\over r^2}-2\s^2A_0'^2r^2\)e^{-8\f}+\s^2\Big(4\D+4r\D''-24r\D\f'-6r^2\D'\f'\Big)e^{-2\f}\bigg]\Bigg]\f''\cr
&+\Bigg[-6(2\s r\D+\s r^2\D'+\s'r^2\D)+\a\bigg[{16P^2\over r^3}(\s r\D'-2\s\D+\s'\D)e^{-8\f}+2\s^2A_0'^2(4\s r\D+\s r^2\D'\cr&-3\s'r^2\D)e^{-8\f}
+\Big(-4\s\D'-4\s'\D+\s^3(12\D\D'+4r\D'^2+4r\D\D'')+3\s^2\s'(4\D^2+4r\D\D')\Big)e^{-2\f}\bigg]\Bigg]\f'\cr&
+\a\s\bigg[\D e^{-2\f}-8\D\(8{P^2\over r^2}-A_0^2r^2\)e^{-8\f}+\s^2(-3r^2\D\D''-3r^2\D'^2-34r\D\D'-16\D^2)e^{-2\f}-9\s\s'r(4\D^2\cr&+r\D\D')e^{-2\f}\bigg]\f'^2
+4\a\s^3r\D(4\D+r\D')e^{-2\f}\f'^3+\Bigg[-6\({P^2\over\s r^2}-\s A_0'^2r^2\)e^{-6\f}+\a\bigg[\Big((\s\D''+\s^3(\D\D''\cr&+\D'^2)
+\s'(-\D'+3\s^2\D\D')\Big)e^{-2\f}+\Big(8A_0'^2(1+\s^2\D)-4A_0'A_0''\s^3\D+\s{P^2\over r^4}\(-6\D+2r\D'+4r^2\D''\)\cr&+2\s'{P^2\over r^3}(\D+2r\D')\Big)e^{-8\f}
+42\s{A_0'^2P^2\over r^2}e^{-14\f}\bigg]\Bigg]=0&(9)
}$$

$$\Bigg[A_0\Bigg\s\bigg[\D'R'R+\D R'^2-3\D R^2\f'^2+A_0'^2R^2e^{-6\f}\bigg]+\a\s\bigg[-3{P^2A_0'^2\over R^2}e^{-14\f}\cr
&-(\D'\f'+2\D\f'^2)e^{-2\f}+\(A_0'^2+{P^2\over R^3}\big(\D'R'+4\D'R\f'+8\D R\f'^2+2\D R'\f'\big)\)e^{-8\f}\bigg]\cr
&+\a\s^3\D\bigg[\(\D'R'^2\f'+2\D'RR'\f'^2+2\D R'^2\f'^2-\D'R^2\f'^3-4\D RR'\f'^3\)e^{-2\f}-A_0'^2(R'^2+R^2\f'^2)e^{-8\f}\bigg]}$$

\end

$$\Bigg[A_0'\[R^2e^{-6\f}+\a\((1-\D R'^2-\D R^2\f'^2)e^{-8\f}-3{P^2\over R^2}e^{-14\f}\)\]\Bigg]'=0\eqno(7)$$
$$ $$
$$\eqalign{&\bigg[6\D R^2\f'+\a\Big[\big(\D'+4\D\f'-\D\D'R'^2-4\D\D'RR'\f'-4\D^2R'^2\f'+3\D\D'R^2\f'^2+12\D^2RR'\f'^2\big)e^{-2\f}\cr
&+\(2A_0'^2\D R^2\f'-{P^2\over R^3}\big(4\D'R+16\D R\f'+2\D R'\big)\)e^{-8\f}\Big]\bigg]'=6\(A_0'^2R^2-{P^2\over R^2}\)e^{-6\f}-\a\bigg[42{P^2A_0'^2\over R^2}e^{-14\f}\cr&
+2\big(\D'\f'+2\D\f'^2+\D\D'R'^2\f'+2\D\D'RR'\f'^2+2\D^2R'^2\f'^2-\D\D'R^2\f'^3-4\D^2RR'\f'^3\big)e^{-2\f}\cr&
-8\(A_0'^2\big(1-\D R'^2-\D R^2\f'^2\big)+{P^2\over R^3}\big(\D'R'+4\D'R\f'+8\D R\f'^2+2\D R'\f'\big)\)e^{-8\f}\bigg]}$$
$$ $$
$$\eqalign{&\Bigg[RR'+\a\[\(-\f'+\D R'^2\f'+2\D RR'\f'^2-\D R^2\f'^3\)e^{-2\f}+{P^2\over R^3}\(R'+4R\f'\)e^{-8\f}\]\Bigg]'\cr
&=R'^2-3R^2\f'^2+\a\bigg[\big(-2\f'^2+\D'R'^2\f'+2\D'RR'\f'^2+4\D R'^2\f'^2-\D'R^2\f'^3\cr&-8\D RR'\f'^3\big)e^{-2\f}
+\(A_0'^2\big(R'^2+R^2\f'^2\big)-{P^2\over R^3}\big(8R\f'^2+2R'\f'\big)\)e^{-8\f}\bigg]}$$
$$ $$
$$\eqalign{&\D'R'R+\D R'^2-1-3\D R^2\f'^2+\(A_0'^2R^2+{P^2\over R^2}\)e^{-6\f}-\a\bigg[3{P^2A_0'^2\over R^2}e^{-14\f}
+\bigg(\D'\f'\cr&+2\D\f'^2-3\(\D\D'R'^2\f'+2\D\D'RR'\f'^2+2\D^2R'^2\f'^2-\D\D'R^2\f'^3-4\D^2RR'\f'^3\)\bigg)e^{-2\f}\cr
&-\(A_0'^2\big(1-3R'^2-3R^2\f'^2\big)+{P^2\over R^3}\big(\D'R'+4\D'R\f'+8\D R\f'^2+2\D R'\f'\big)\)e^{-8\f}\bigg]=0}$$

From (7) immediately follows that
$$A_0'={QR^2e^{6\f}\over R^4+\a\Big(R^2(1-\D R'^2-\D R^2\f'^2)e^{-2\f}-3P^2e^{-8\f}\Big)}\eqno(11)$$

\end